\documentclass{JHEP3}

\usepackage{amssymb}

\usepackage{amsmath,amssymb,graphics}
\usepackage{epsfig,multicol}
\usepackage{graphicx}
\usepackage{epsfig,subfigure}
\def\L{{\Omega_\Lambda}}    \def\M{{\Omega_m}}       \def\K{{\Omega_k}}

\title{Fitting cosmological data to the function $q(z)$ from GR Theory: Modified Chaplygin Gas}
\author{Alan M. Velasquez-Toribio\thanks{alan@fisica.ufjf.br} and Maria Luiza Bedran \thanks{On leave from Universidade Federal do Rio de Janeiro. Email:bedran@fisica.ufjf.br}\\
Departamento de F\'{\i}sica - ICE - \\Universidade Federal de Juiz de Fora, Brazil.\\
CEP: 36036-330, MG,  Brazil}

\abstract{
In the Friedmann cosmology the deceleration of the expansion $q$ plays a fundamental role. We derive the deceleration as a function of redshift $q(z)$ in two scenarios: $\Lambda$CDM model and modified Chaplygin gas ($MCG$) model. The  function for the $MCG$ model is then fitted to the cosmological data in order to obtain the cosmological parameters that minimize $\chi^2$.
We use the Fisher matrix  to construct the covariance matrix of our parameters and reconstruct the q(z) function. We use Supernovae Ia, WMAP5 and BAO measurements to obtain the observational constraints. We determined  the present acceleration as $q_0=-0.60 \pm 0.12$ for the $MCG$ model using the Constitution dataset of SNeIa and BAO, and $q_0=-0.63 \pm 0.17$ for the Union dataset and BAO. The transition redshift from deceleration to acceleration was found to be around $0.6$ for both datasets. We have also determined the dark energy parameter for the $MCG$ model: $\Omega_{X0}=0.834 \pm 0.028$ for the Constitution dataset and $\Omega_{X0}=0.854 \pm 0.036$ using the Union dataset.}

\begin{document}

\section{Introduction}
During the last decade, the observation of type Ia supernovae
(SNeIa) and the cosmic microwave background radiation (CMBR) permitted the
determination of the cosmological parameters with ever increasing
precision. The reported results of the seven year analysis of WMAP
\cite{Larson} are $\Omega_{\Lambda}=0.734 \pm 0.029$ and $\Omega_{m}=0.266 \pm
0.029$; these values were obtained assuming a flat geometry
($\Omega_{k0}=0$). From measured luminosity distances to SNeIa, Riess et
al.\cite{Riess,Riess1} determined the redshift of the transition from
decelerated to accerelerated expansion to be $z_t=0.46 \pm 0.13$;
this value was obtained assuming a linear expansion for the
deceleration parameter, that is,~ $q(z)=q_0+q_1 z$. The problem with this 
linear expansion is that it works well for small redshifts but the transition 
redshift is not so small. Using another parametrization for $q(z)$  
Shapiro and Turner \cite{Shapiro} concluded that the present SNeIa data 
cannot rule out the possibility that the universe has been decelerating 
since $z=0.3$. In other references (\cite{nesseris},\cite{Cunha}) 
it was shown that the value of the transition redshift depends on the adopted parametrization for $q(z)$, 
as well as on the data sample. 
In particular, a parametrization that has the redshift transition
as free parameter has been presented in reference \cite{ishida}. This parametrization could be used to study the
kinematics of the expansion regardless of the matter content of the Universe.
However, the statistical properties of data are still not good enough 
to produce strong constraints.  
In general, the parametrizations are helpful by their 
phenomenological properties, since they can serve to study the accelerated expansion
in different contexts, such as the structure formation, that would be difficult 
to study into a fundamental theory.

Among all possible candidates to explain the accelerated expansion, the Chaplygin gas is a strong candidate;
it is the best known proposal of a unification of dark matter ($DM$) with dark energy ($DE$) into a single fluid.
The ideia is: an equation of state ($EoS$) leads to a component which behaves as dust at early stage and as cosmological constant at later stage. Following this idea it was considered the so called generalized Chaplygin gas \cite{BBS}.
This model has been analyzed many times in the literature; see for example \cite{mg}. Subsequently  this model has been modified to include an initial phase of radiation, and is called the modified Chaplygin gas ($MCG$). From the theoretical point of view, this scenario can also be restated as a Friedmann-Lema\`{i}tre-Robertson-Walker ($FLRW$) cosmological model containing
a scalar field $\phi$ with its self-interacting potential \cite{mcg}.  
 
In this paper we have as main aim to study the $MCG$ model using several data to constraint as much as possible the parameters of the model. In the literature the generalized Chaplygin gas has been studied in the context of statefinder diagnostic, stochastic gravitational waves, observational constraints using  gamma ray bursts, strong lensing, Supernovae Ia, etc.; see \cite{mgas} for references. We studied the deceleration parameter in the $MCG$ and used recent observational data to constraint its free parameters.  
We used two sets of data of type Ia Supernovae: the Constitution set and the Union set. We also use the CMB and BAO data. In order to make comparisons, we derive the function $q(z)$ for the standard cosmological model, that is the $\Lambda CDM$ model. The derivation was performed with the minimal assumptions: GR theory is valid and the universe is homogeneous and  isotropic, that is, the cosmological metric is the $FLRW$ one.  No assumption was made about the spatial curvature of the universe, however the analysis on the observational limits of the free parameters was restricted to flat models. We leave for a future paper the models with curvature.

The outline of this paper is as follows. In section 2 we present the equations of the models, $\Lambda CDM $ and $MCG$, and determine the Hubble parameter and the deceleration parameter respectively. In section 3
we describe the method used to obtain the confidence regions and the reconstruction of the function $q(z)$. Finally,  section 4 is devoted to the discussion of our results.

\section{Equations of our Models}

\subsection{$\Lambda CDM$ Model}
Let's assume that the universe is described by the FRW metric and the energy content is a pressureless fluid and a cosmological constant. The first Friedmann equation for the scale factor $a(t)$ reads

\begin{equation}
H^2=\frac{\dot a ^2}{a^2}=\frac{8\pi G\rho}{3}+\frac{\Lambda}{3}-\frac{k}{a^2}
\label{Friedmann}
\end{equation}

\noindent which can be written as

\begin{equation}
\M+\L+\K=1                            \label{Friedmann1}
\end{equation}

\noindent with the definitions

$$ \M = \frac{8\pi G \rho}{3H^2}~~~~~~~~\L=\frac{\Lambda}{3H^2} ~~~~~~~~\K=\frac{-k}{a^2H^2}$$

The Bianchi identity for a pressureless fluid can be integrated to give

\begin{equation}
\rho a^3 = \rho_0 a_0^3                          \label{Bianchi}
\end{equation}

\noindent where the subscript zero denotes present values of the quantities. Inserting eq.(\ref{Bianchi}) into eq.(\ref{Friedmann}) and writing $-k$ as 

$$ -k = H^2 a^2 \K = H_0^2 a_0^2 \K_0  $$

\noindent we obtain

\begin{equation}
{\dot a}^2= H_0^2 a_0^2 \left[ \M_0~\frac{a_0}{a} + \L_0~\frac{a^2}{a_0^2} + \K_0 \right]
\label{eq.4}
\end{equation}
\vspace{.4cm}

In terms of the redshift $z=\frac{a_0}{a}-1$, eq.(2.4) reads

\begin{equation}
\dot a^2 = \frac{H_0^2 a_0^2}{(1+z)^2}~E^2(z)= H_0^2 a^2(z) E^2(z)      \label{eq.5}
\end{equation}

\noindent where

\begin{equation}
E^2(z)=\M_0~(1+z)^3 + \K_0~(1+z)^2 + \L_0                \label{Q(z)}
\end{equation}
\vspace{.4cm}

Using eq.(2.5) we can relate $\frac{da}{dz}$ with $\dot a =\frac{da}{dt}$:

\begin{equation}
\frac{da}{dz}=-\frac{a_0}{(1+z)^2}=-\frac{a(z)}{1+z}=-\frac{1}{H_0 (1+z) E(z)}~\frac{da}{dt}                                                \label{da/dz}
\end{equation}

\noindent from which we infer the relation between $dt$ and $dz$:

\begin{equation}
\frac{dz}{dt}= - H_0 (1+z) E(z)                \label{dz/dt}
\end{equation}

Now, using eq.(\ref{da/dz}) and calculating $\ddot a$, we obtain for the deceleration parameter

$$q = - \frac{ \ddot a}{H^2 a}             \label{q}$$

the expression    

\begin{equation}
q(z)= \left[\frac{\M_0}{2}(1+z)^3-\L_0\right]\left[\M_0(1+z)^3+\K_0(1+z)^2+\L_0\right]^{-1}
\label{q(z)}
\end{equation}  \vspace{.5cm}

\noindent For $z>>1$ we see that $q(z)=0.5$ as expected. The transition redshift, where $q(z_t)=0$, is given by 

$$ (1+z_t)^3= \frac{2 \Omega_{\Lambda 0}}{\Omega_{m0}} $$

\noindent for any value of the curvature $k$. 

In Figure 1 we reconstruct the evolution of q(z) using the Constitution dataset and Union dataset. From this figure we find that both datasets predict similar transition redshifts, $z_{t} \approx 0.5$. Our reconstruction was made with a $1 \sigma$ confidence level.

\begin{figure}[htb]
\begin{center}
\includegraphics[height= 7.0 cm,width=7.5cm]{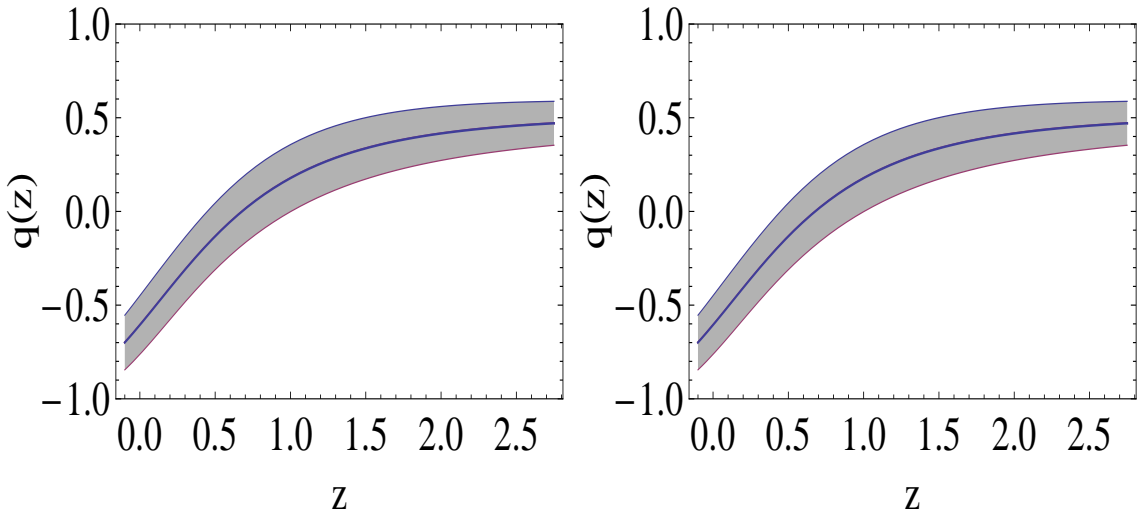}
\includegraphics[height= 7.0 cm,width=7.5cm]{LCDMconstqz.eps}
\end{center}
\caption{We show the observational constraints for the $\Lambda CDM$ model using only SNeIa with $\Omega_{b0}=0.042 \pm 0.027$. Left panel: Constitution set; right panel: Union set.}
\label{fig1}
\end{figure}

\subsection{Modified Chaplygin gas (MCG) model without cosmological constant}
  
    The generalized Chaplygin gas model has been proposed as a source term in Einstein's field equations in order to unify the concepts of cold dark matter and dark energy \cite{Kamenshchik,Bilic,Bento}. We will consider the energy content of the universe as a fluid that behaves like a perfect fluid of non-zero pressure at early times, like a pressureless fluid at intermediate times and like dark energy at present.
  
The equation of state of the MCG is given by \cite{mcg}
  
\begin{equation}
p = ~B\rho - \frac{A}{{\rho}^{\alpha}}           \label{eq.state} \,,
\end{equation}

\noindent where $A$,$B$ and $\alpha$ are non-negative constants. For $B=0$ we have the pure generalized Chaplygin gas and for $A=0$ a perfect fluid. The MCG behaves as radiation (when $B = 1/3$) or dust-like matter (when $B = 0$) at early stage, while as a cosmological constant at later stage.
On the other hand, the Bianchi identity

\begin{equation}
3\,{\frac {\dot a} a}\,(p + \rho) + \dot {\rho} =0 \,
\end{equation}

\noindent yields after integration (see \cite{Bedran} for details):

\begin{equation}
\rho (a) = \rho _0 \left[ {\Omega _X  + \left( {1 - \Omega _X } \right)\left( {\frac{{a_0 }}{a}} \right)^{3R} } \right]^{{1 \mathord{\left/
 {\vphantom {1 {\left( {\alpha  + 1} \right)}}} \right.
 \kern-\nulldelimiterspace} {\left( {\alpha  + 1} \right)}}}  \,,
\end{equation}

\noindent where $$R=(B+1)(\alpha +1)$$ 
$\rho_0$ is the present energy density,  and we define the dimensionless parameter

\begin{equation}
\Omega_X = \frac{A}{(B+1)\,\rho_0^{\alpha+1}} \,.         \label{OmegaX}
\end{equation}

The same analysis of the previous section can be done, leading to the following expression for the deceleration parameter:

$$
q(z)=\frac{E^{-2}(z)}{2} \left[\Omega_X+(1-\Omega_X)(1+z)^{3R}\right]^{-\alpha/(1+\alpha)}$$
\begin{equation}
\left\{(1-\Omega_X)(3B+1)(1+z)^{3R} - 2\Omega_X\right\}                 \label{q(z)MCG}
\end{equation}

\noindent where

\begin{equation}
E^2(z)= [\Omega_{X} + (1-\Omega_{X})(1+z)^{3R}]^{1/(1+\alpha)}+\Omega_{k0}(1+z)^{2}
\end{equation}

The transition redshift for the MCG model is given by:

\begin{equation}
(1+z_t)^{3R}=\frac{2\Omega_X}{(3B+1)(1-\Omega_X)}.   \label{ztMCG}
\end{equation}

The dimensionless parameter $\Omega_X$ represents the fraction of dark energy in the content of the universe, thus taking the value $\Omega_X \approx 0.7$. If we chose $B=1/3$ in order to describe the evolution of the universe since the radiation era, and consider $0 <\alpha < 0.5$, which is required from thermodynamical considerations (see \cite{Santos}), we find $z_t \approx 0.2$. If we consider a pure Chaplygin gas $(B=0)$ and  $\Omega_X \approx 0.7$, the value $z_t=0.46$ can be achieved if $\alpha \approx 0.4$. This result is compatible with the analysis done in \cite{Makler} for a Chaplygin gas in the flat ($k=0$) case.

Now, in order to carry out our analysis of observational constraints, we consider as components of the Universe: baryons plus $MCG$. Thus, in the flat case, the Hubble parameter is given by:

\begin{equation}
E^2(z) = \left(\frac{H(z)}{H_{0}}\right)^{2}= \Omega_{b0}(1+z)^3 + (1-\Omega_{b0})[(1-\Omega_{X0})(1+z)^{3R}+\Omega_{X0})]^{1/(1+\alpha)}
\end{equation}

This expresion will be used in our analysis of observational constraints in the following section. As can be seen, we have four free parameters $(\Omega_{b0},\Omega_{X0},\alpha,B)$.

\section{Some Observational Constraints}

In the present section we consider some observational constraints of SNeIa, BAO and CMB for our models.
In this context it is important to consider the comoving distance to an object at redshift $z$,

\begin{equation}
r(z) = cH_{0}^{-1}\int{\frac{dz'}{E(z')}},
\end{equation}

\noindent where we consider only the flat case; $E(z)=H(z)/H_{0}$ is given for the $\Lambda CDM$ model by equation (2.6) and for the $MCG$ model by equation (2.17). Using the equation above, the luminosity distance in the flat case is then given by $d_{L}=(1+z)r(z)$.

\subsection{Constraints from Supernovae Data}

The supernovae Ia data give us  the distance modulus ($\mu$) to each supernova, that is given by
\begin{equation}
\mu \equiv m_{obs}(z_{i})-M = 5 \log[\frac{d_{L}}{Mpc}]+25
\end{equation}
where $M$ is their absolute magnitudes. The distance modulus also can be written as
\begin{equation}
\mu=5\log_{10}D_{L}(z)+\mu_{0}
\end{equation}
where $D_{L}=\frac{H_{0}d_{L}}{c}$ is the Hubble-free luminosity distance and $\mu_{0}$ is the zero point offset (which is an additional model-independent parameter) defined by 
\begin{equation}
\mu_{0}=5\log_{10}(\frac{cH_{0}^{-1}}{Mpc}) +25 = 42.38-5\log_{10} h
\end{equation}

In the present paper we used the Union set incluing 307 data of Kowalski et. al \cite{kowalski}, that includes the recent samples from the SNLS \cite{snls} and ESSENCE Surveys \cite{essence}, older datasets, as well as the recently extended dataset of distant supernovae observed with HST \cite{hst}. The sample consisting of 414 SNeIA was reduced to 307 data after various selection cuts were applied in order to create a homogeneous sample.
We also used the so called "Constitution" set of Hicken et. al \cite{hicken} incluing 397 data, out of which 100 come from the new low-z CfA3 sample and the rest from the Union set. Both samples have a redshift range of $0.015 \leq z \leq 1.55$. The main improvement of the Constitution sample is the inclusion of a larger number of nearby $(z<0.2)$ SNeIa; their inclusion helps to reduce the statistical uncertainty \cite{hicken}.

The statistic $\chi^{2}$ is a useful tool for estimating goodness-of-fit and confidence regions on parameters. In our case the $\chi^{2}_{SN Ia}$ is given by 
\begin{equation}
\chi^{2}_{SN Ia}(p_{i}) = \sum_{i=1}^{n}{\frac{(\mu_{the}(p_{i},z_{i})-\mu_{obs,i}(z_{i}))^{2}}{\sigma_{obs,i}^{2}}}
\end{equation}
 
\noindent where $p_{i}=(\Omega_{b0},\Omega_{X0},b,B)$. The $\chi^{2}$ function can be minimized with respect to the $\mu_{0}$ parameter, as it is independent
of the data points and the dataset. Expanding the equation above with respect to $\mu_{0}$, we obtain:

\begin{equation}
\chi^{2}(p_{i})_{SN Ia} = A(p_{i})-2\mu_{0}B(p_{i})+\mu_{0}^{2}C(p_{i})
\end{equation}

\noindent which has a minimum for $\mu_{0} = B(p_{i})/Cp_{i})$, giving
\begin{equation}
\chi^{2}_{SN Ia,min} = \bar{\chi}^{2}_{SN Ia} = A(p_{i}) - \frac{B^{2}(p_{i})}{C(p_{i})}
\end{equation}

where
\begin{eqnarray}
A(p_{i})&=&\sum_{i}^{n}{\frac{(\mu_{th} - \mu_{obs}(p_{i},\mu_{0}=0))}{\sigma_{i}}^{2}}\\
B(p_{i}) &=& \sum_{i}^{n}{\frac{\mu_{th} - \mu_{obs}(p_{i},\mu_{0}=0)}{\sigma_{i}}}\\
C(p_{i})& =& \frac{1}{\sigma_{i}^{2}}
\end{eqnarray}

Now this new $\bar{\chi}^{2}_{SN Ia}$ is independent of $\mu_{0}$ and can be minimized with respect to the parameters
of the theoretical model.

\subsection{CMB}
In our investigation we used the method proposed by Komatsu. et al. \cite{Komatsu}; in this case two distances are important. The distance to recombination can be written as:
\begin{equation}
R=\sqrt{\Omega_{m0}}r(z_{CMB})
\end{equation}
where $r(z_{CMB})$ is given by equation (3.1). The second distance is the angular scale of the sound horizon at recombination, $r_{s}$, that allows to determine the first acoustic peak in the following form:

\begin{equation}
l_{a} = \pi \frac{r(z_{CMB})}{r_{s}(z_{CMB})}
\end{equation}

\noindent where $r_{s}$ is defined as follows:

\begin{equation}
r_{s}(a_{CMB}) = \frac{1}{H_{0}}\int_{0}^{a_{CMB}}{\frac{c_{s}(a)}{a^{2}E(a)}da}
\end{equation}
where $c_{s}$ is the sound speed and is given by $c_{s}=1/\sqrt{3(1+\frac{3 a \Omega_{b0}}{4\Omega_{\gamma}})}$. We are considering $z_{CMB} = 1090$ \cite{Komatsu}. Following this reference, we used the prescription for the WMAP5 distance priors. Thus the $\chi^{2}$ for the CMB data is:

\begin{equation}
\chi^{2}_{CMB}= X^{T}_{CMB}C^{-1}_{CMB}X_{CMB}
\end{equation}

\noindent where $X^{T}$ denote its transpose and the vector $X=(R,l_{A},100 h^{2} \Omega_{b0})$ and $C^{-1}$ is the inverse covariance matrix, which is given for the WMAP5 observations by:

\begin{equation}
\mathbf{C_{i,j}}=\left(\begin{array}{ccc}
0.000367364 &  0.00181498 &  -0.000201759\\
0.00181498 & 0.731444 & -0.0315874 \\
-0.000201759 & -0.0315874 &0.00355323\\
\end{array}\right)
\end{equation}

The best fit turns out to be:

\begin{equation}
\mathbf{X_{CMB}}=\left(\begin{array}{ccc}
R-1.70\\
l_{a}-302.10 \\
100h^{2} \Omega_{b0}-2.2765\\
\end{array}\right)
\end{equation}

\subsection{BAO}

We used BAO distance measurements obtained at $z = 0.20$ and $z = 0.35$ from joint analysis of the 2dFGRS and SDSS data \cite{baodata}. The distance scale used is a combination between the radial dilation and the square of the transverse dilation; this is
 
\begin{equation}
D_{V}(z_{BAO}) = \left[\left(\int_{0}^{BAO}{\frac{dz}{H(z)}}\right)^{2}\frac{z_{BAO}}{H(z_{BAO})}\right]^{1/3}
\end{equation}

We also apply the maximum likelihood method and in this case the $\chi^{2}$ is given by:

\begin{equation}
\chi_{BAO}^{2}= X_{BAO}^{T} C^{-1}_{BAO} X_{BAO}
\end{equation} 
where the inverse covariance matrix is
\begin{equation}
\mathbf{C_{BAO}^{-1}}=\left(\begin{array}{cc}
35059&  -24031 \\
-24031 & 108300 \\
\end{array}\right)
\end{equation}
and

\begin{equation}
\mathbf{X_{BAO}}=\left(\begin{array}{ccc}
\frac{r_{s}(z_{des})}{D_{V}(0.20)}-0.1980\\
\frac{r_{s}(z_{des})}{D_{V}(0.35)}-0.1094\\
\end{array}\right)
\end{equation}
where $r_{s}$ is given by equation (3.13).

\subsection{Combining the Datasets}

We considered that the observational data are independent, so we defined the $\chi^{2}_{total}$ as
\begin{equation}
\chi^{2}_{total}= \bar{\chi}_{SNIa}^{2} + \chi_{BAO}^{2} + \chi^2_{CMB}
\end{equation}
The best fit values the model can be determined by minimizing the total $\chi^{2}$.
For Gaussian distributed measurements, the $\chi^{2}$ function is directly related to the maximum likelihood estimator. Moreover, if we want to impose a Gaussian prior on one of the parameters being measured, $p_{i}$, centered around $p_{i0}$, with variance $\sigma_{p_{i}}^{2}$, we can use the Bayes theorem, and write the expression
\begin{equation}
L= \exp \left(-\frac{\chi^{2}_{total}}{2}\right)\exp\left[- \frac{(p_{i} - p_{i0})^{2}}{2\sigma_{p_{i}}^{2}}\right]
\end{equation}

In order to constraint the parameters of our interest, we marginalize over the other parameters. To account for the uncertainty of the Hubble parameter, we treat it as a free parameter and then fix it by using the best fit value of the data. For the reconstruction of the q(z) function we used type Ia Supernovae and BAO data. We followed the standard methodology using the Fisher matrix for generating errors. For detais of the method of propagation of errors see references \cite{Alam,heavens}.

\begin{table}\center
\begin{tabular}{|c|c|c|}
\hline
Parameter & Constitution + CMB + BAO & Union + CMB + BAO \\ \hline \hline
B& 0.061 $\pm$ 0.079 & 0.110 $\pm$ 0.097\\ \hline
$\alpha$&0.053 $\pm$ 0.089& 0.089 $\pm$ 0.099 \\ \hline
$\Omega_{X0}$ & 0.834 $\pm$ 0.028& 0.854$\pm$0.036 \\ \hline
$q_{0}$ &-0.60 $\pm$ 0.12&-0.63 $\pm$ 0.17 \\ \hline
$z_{t}$ & 0.62&0.59 \\ \hline
\hline
\end{tabular}
\caption{The best fit parameters of the $MCG$ model. The error bars are obtained by marginalized likelihood analysis that can be obtained from eq.(3.22).}
\end{table}

\begin{figure}[htb]
\begin{center}
\includegraphics[height= 5.0 cm,width=5.5cm]{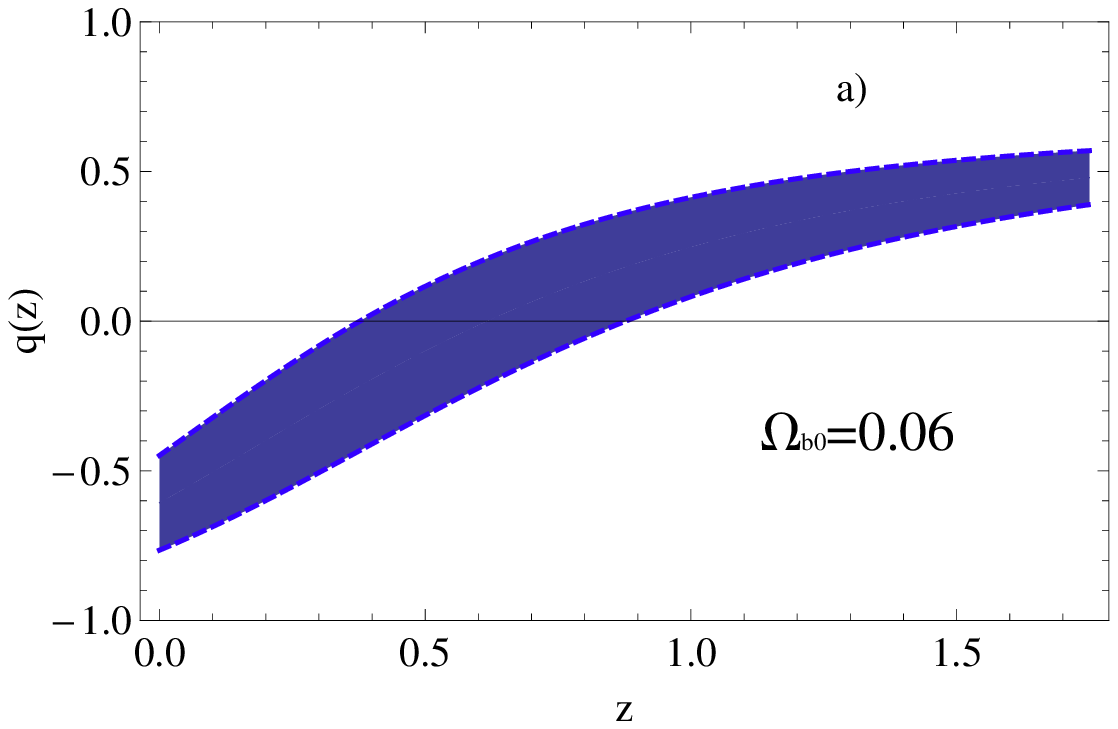}
\includegraphics[height= 5.0 cm,width=5.5cm]{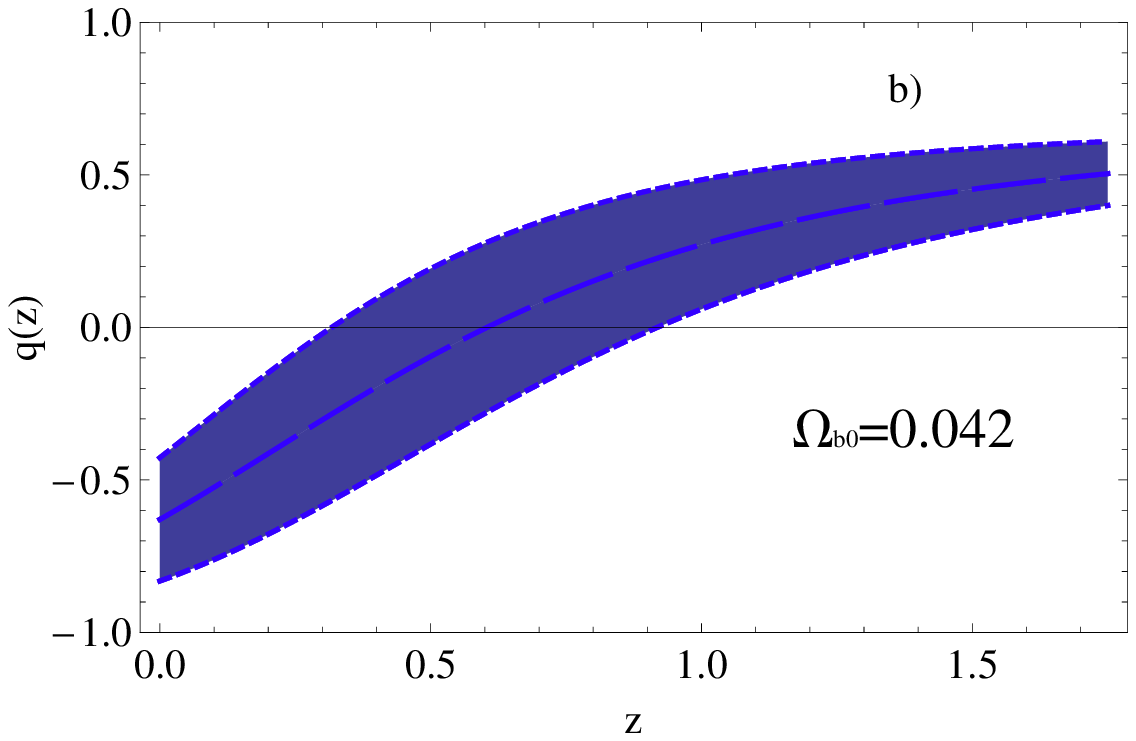}
\includegraphics[height= 5.0 cm,width=5.5cm]{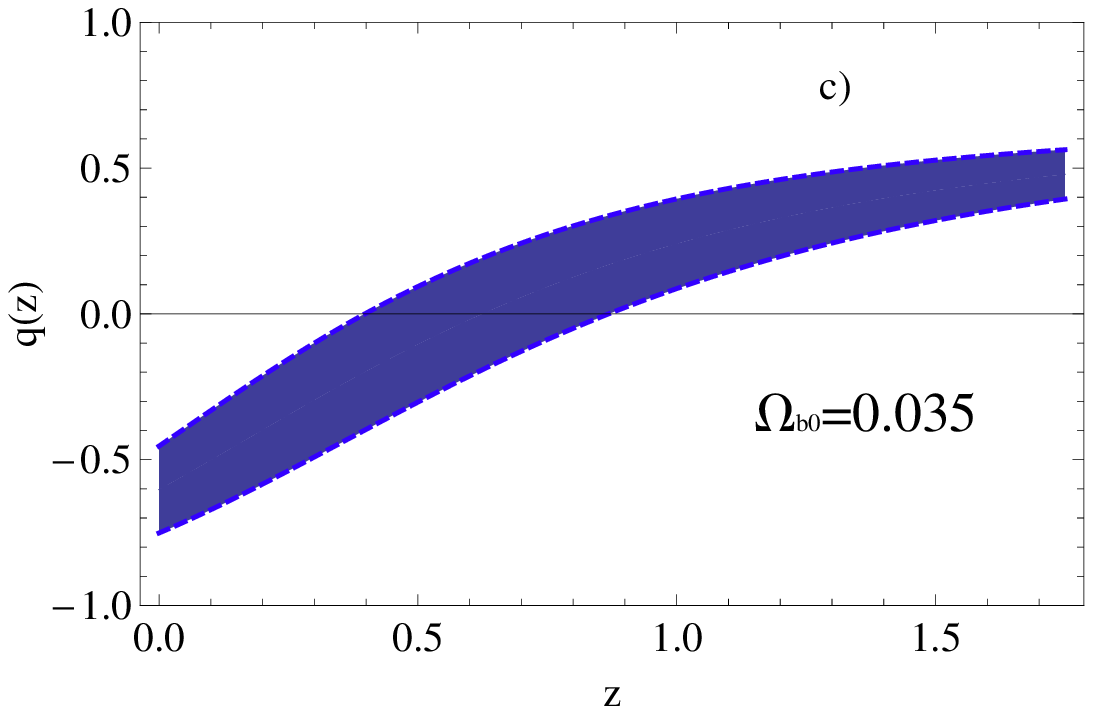}
\includegraphics[height= 5.0 cm,width=5.5cm]{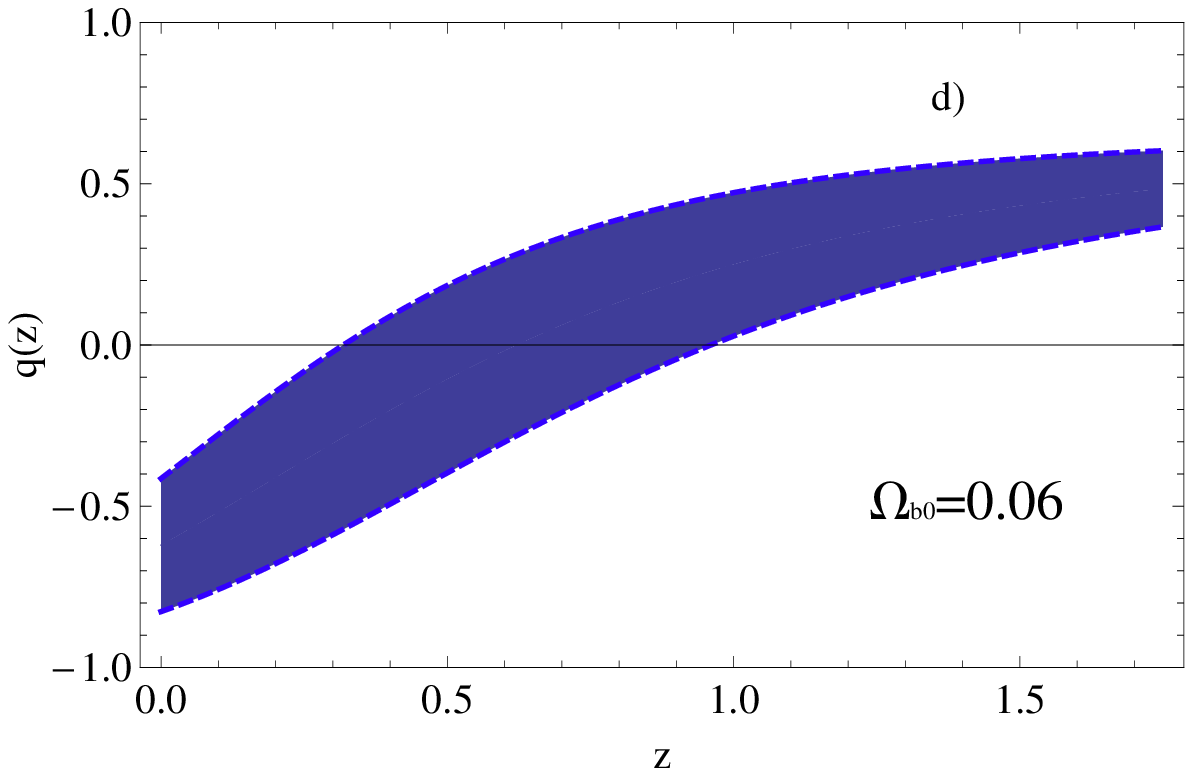}
\includegraphics[height= 5.0 cm,width=5.5cm]{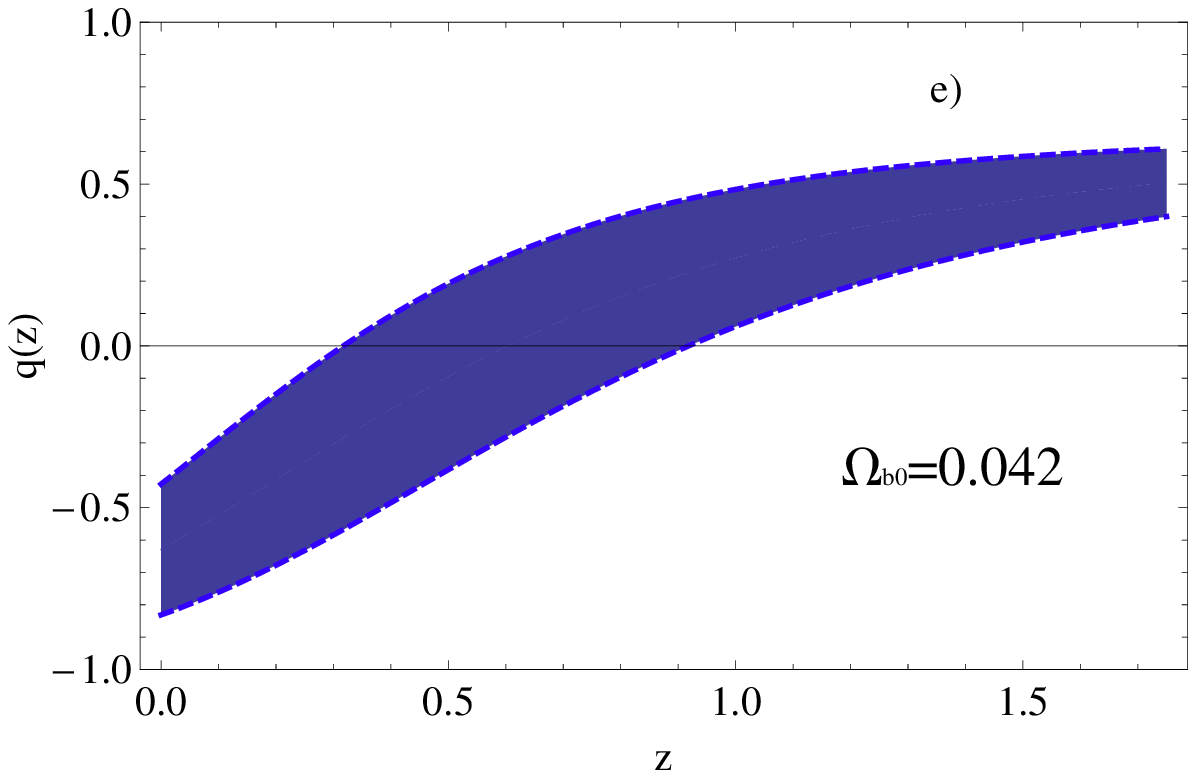}
\includegraphics[height= 5.0 cm,width=5.5cm]{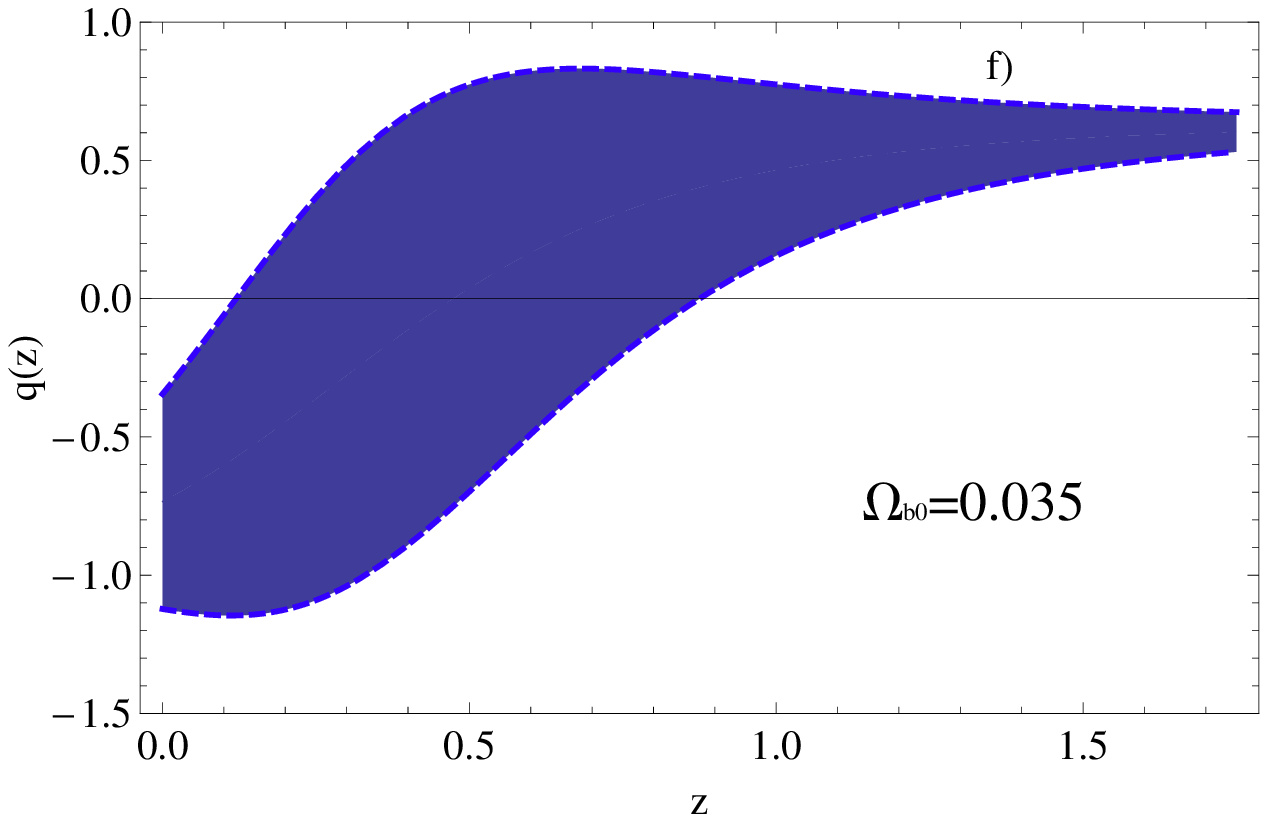}
\end{center}
\caption{In panels a,b,c we present the best fit and reconstruct the erros for the Constitution dataset + BAO, using $B=0.061$, and in panels d,e,f for the Union dataset + BAO, using $B=0.110$.}
\label{fig1}
\end{figure}

\section{Results and Discussion}

The results we obtained for the parameters of the modified Chaplygin equation of state $P=B\rho -A\rho ^{-\alpha}$ are presented in Table 1. We can see that, for both datasets, the exponent $\alpha$ is less than $0.1$. This shows that the negative pressure component $A\rho ^{-\alpha}$ does not differ too much from a cosmological constant. The parameter $B$, related to the dark matter component with positive pressure, is also of order $0.1$ for both datasets. Theoretically, $B$ would be zero for dust-like matter and $1/3$ for radiation. The dark energy parameter $\Omega_{X0}$ was found to be approximatelly $0.8$, a little bit larger than the $0.7$ value of the $\Lambda CDM$ model. Finally, the kinematical parameter $q_0$ is of order $-0.6$ for both datasets, while the transition redshift is $z_t \approx 0.6$. These results for $q_0$ and $z_t$ are consistent with constraints obtained recently with different methods (\cite{Cunha,Lima}).

In Figure 2 we show the behavior of the reconstruction of the $q(z)$ function for the $MGC$ model. In the cases $a$, $b$ and $c$ we used the Constitution dataset + BAO data and in the cases $d$, $e$ and $f$ we used the Union dataset + BAO data. In general the results using the Constitution sample depend less on the $\Omega_{b0}$ value. In the case $f$ we can see a strong change in the reconstruction of the errors of the $q(z)$ function. In Figure 3 we show the one-dimensional likelihood behavior in the case Constitution + BAO + WMAP5 for the $MCG$ model. In the upper panels of Figure 4 we show observational constraints using only type Ia Supernovae data. These results change with the inclusion of the BAO data, but if we include the WMAP5 data we obtain results which strongly reduce the space of parameters. These results are displayed in Figure 5, in which we clearly see that for both samples the $\alpha = 0$ value is included within the confidence region of the space parameters. The $\Lambda CDM$ model is the particular case $\alpha =0$ and $B=0$. Therefore, the points with $\alpha =0$ in Figure 5 do not represent the $\Lambda CDM$ model, because we used the best fit for the $B$ value (see Table 1); the finite value of $B$ is the distinctive feature of the $MCG$ model.

\begin{figure}[htb]
\begin{center}
\includegraphics[height= 4.0 cm,width=4.5cm]{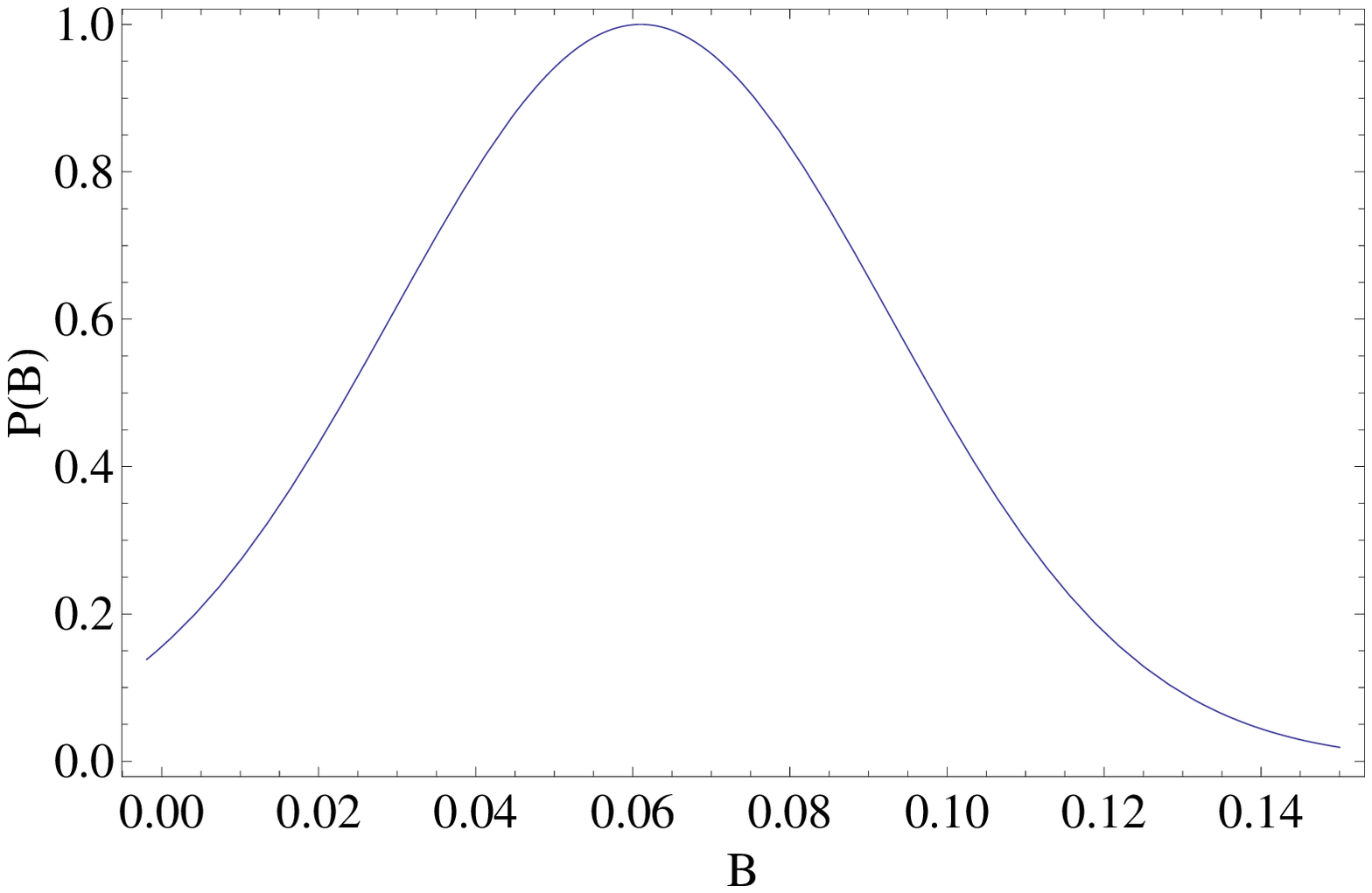}
\includegraphics[height= 4.0 cm,width=4.5cm]{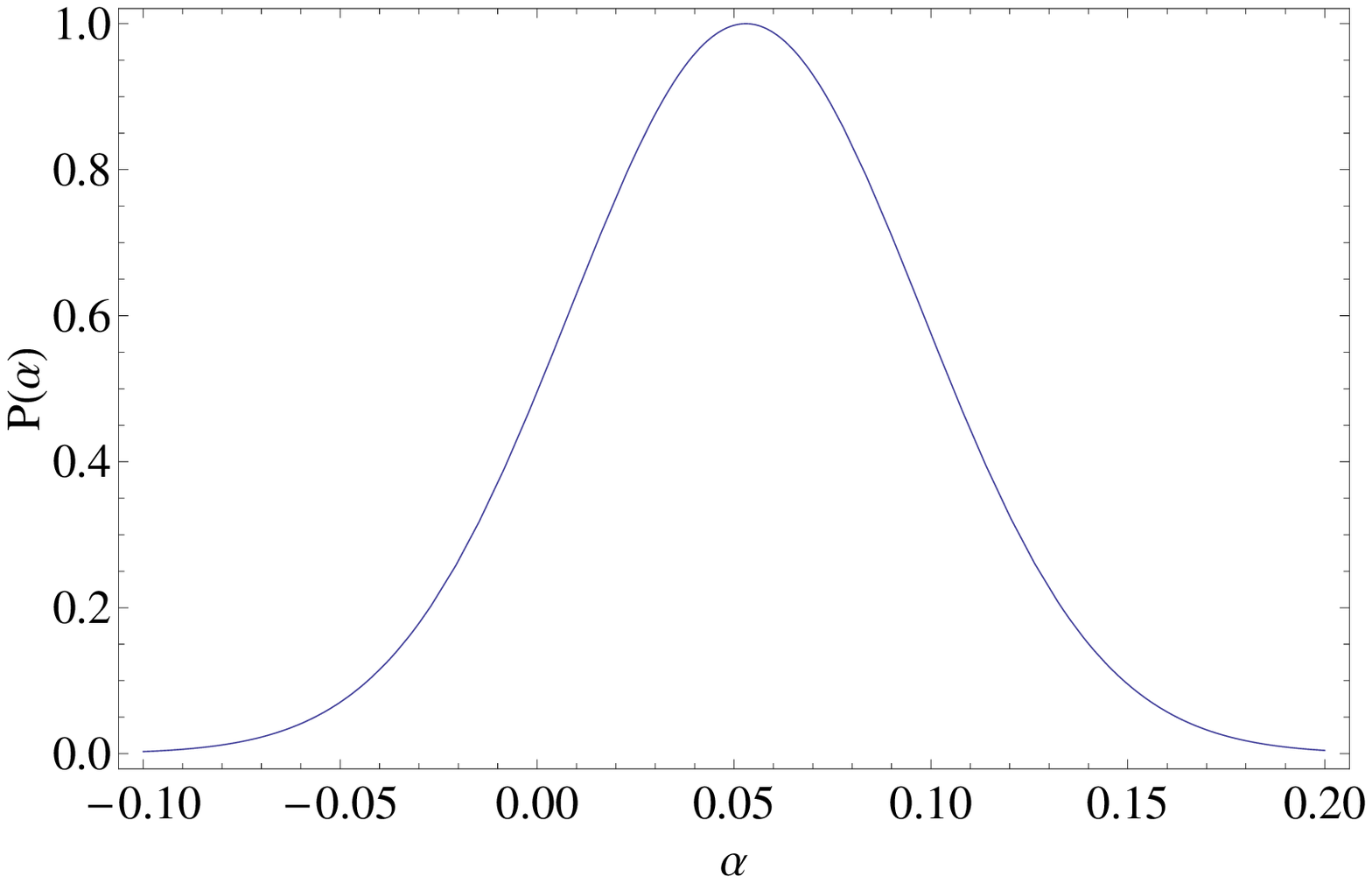}
\includegraphics[height= 4.0 cm,width=4.5cm]{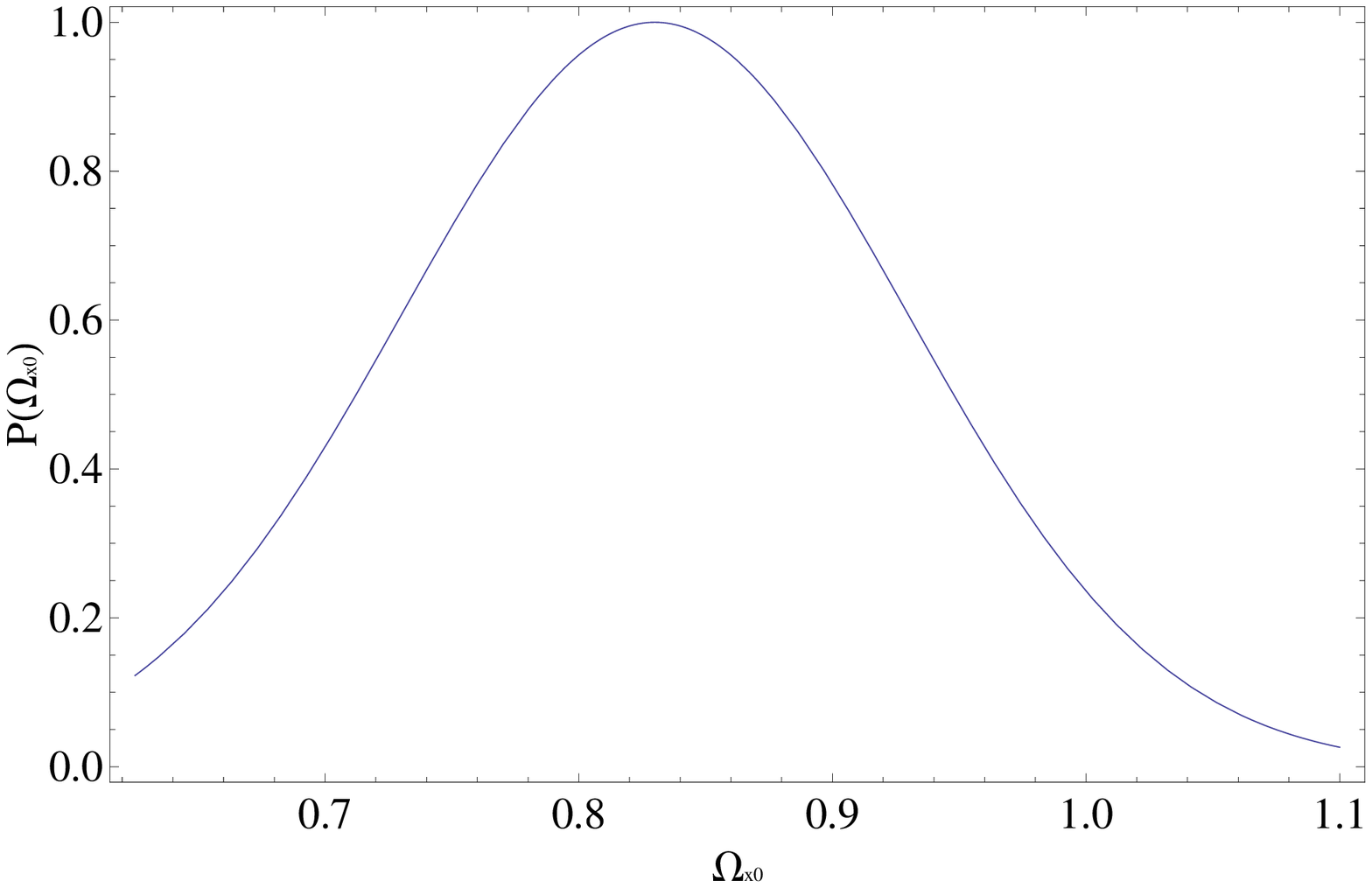}
\end{center}
\caption{The marginalized one-dimensional probabilities likelihood for the parameters of the MCG. The results are for the combined SNeIa (Constitution set), BAO and WMAP5 data. The nuisance parameter $H_{0}$ in the SNeIa is analytically marginalized over.}
\label{fig1}
\end{figure}

\begin{figure}[htb]
\begin{center}
\includegraphics[height= 7.0 cm,width=7.5cm]{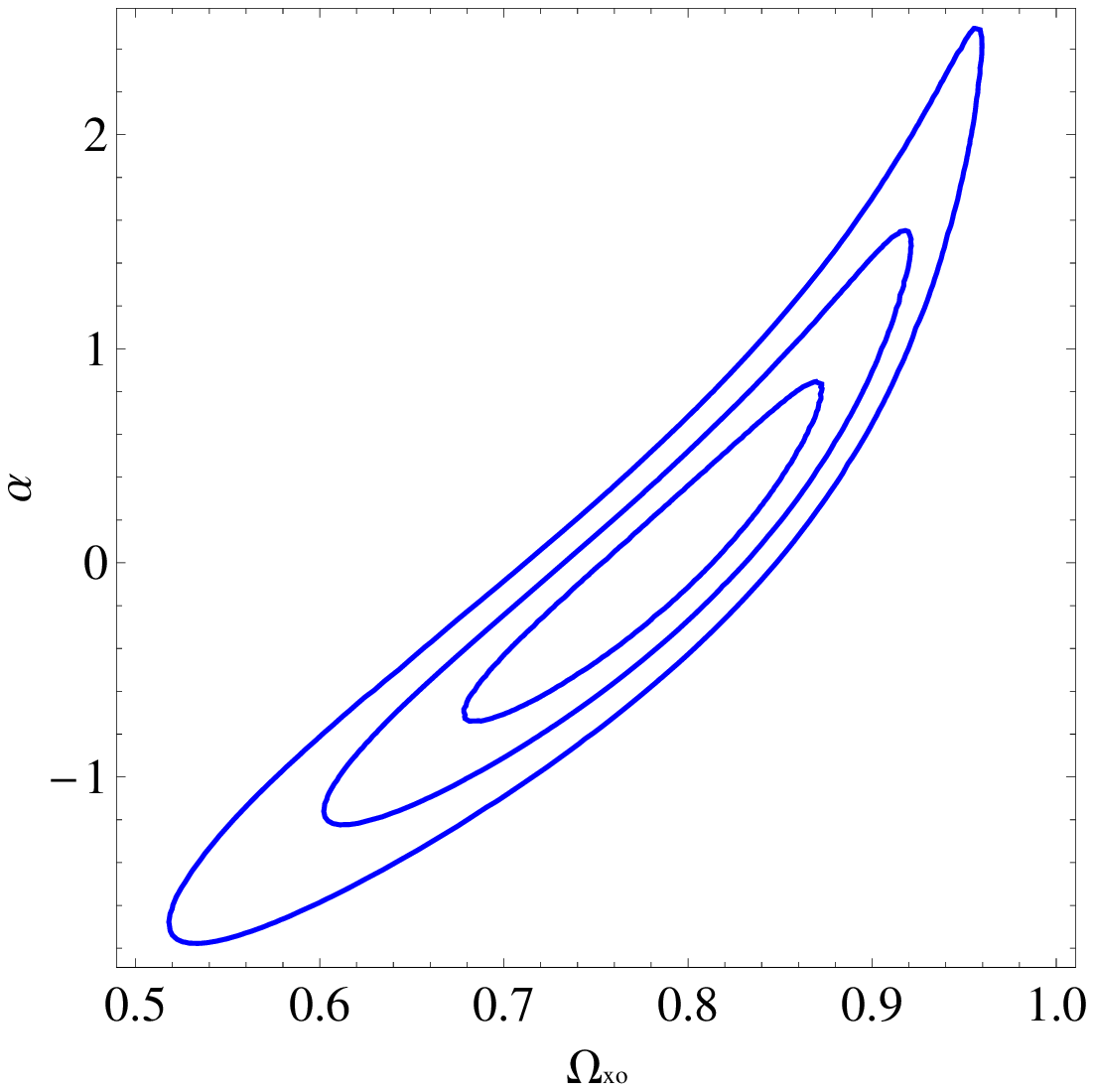}
\includegraphics[height= 7.0 cm,width=7.5cm]{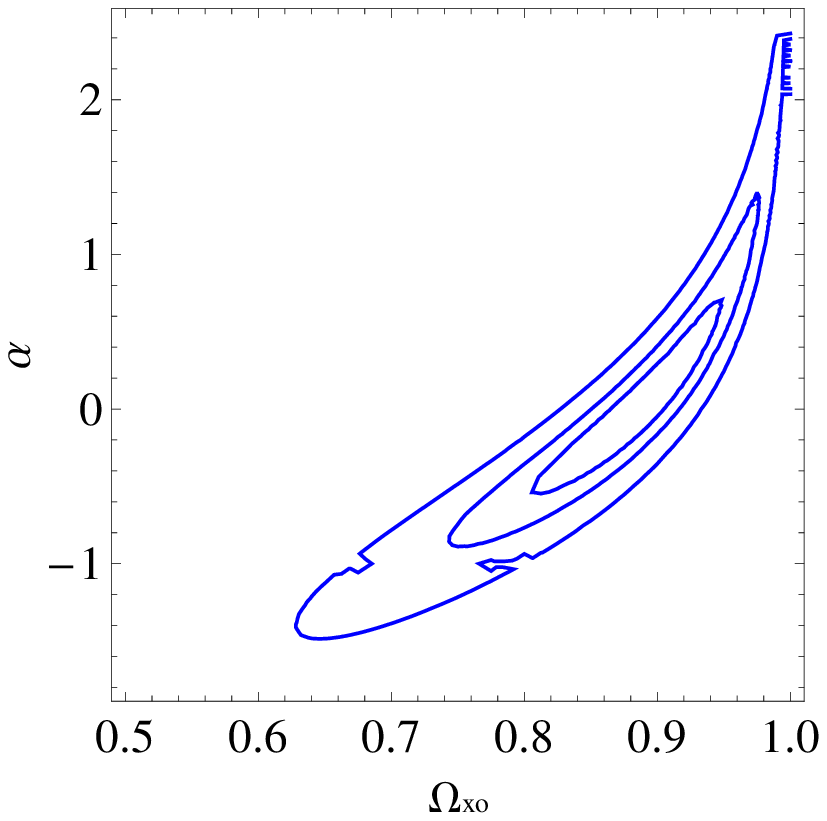}
\includegraphics[height= 7.0 cm,width=7.5cm]{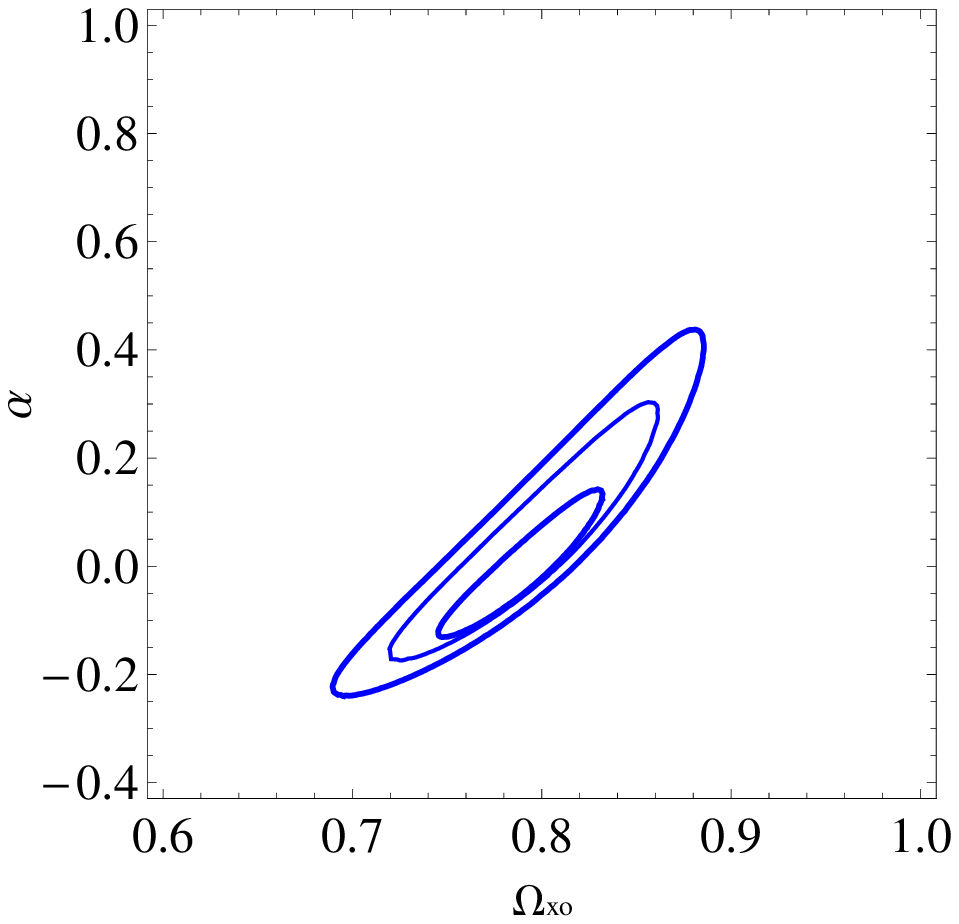}
\includegraphics[height= 7.0 cm,width=7.5cm]{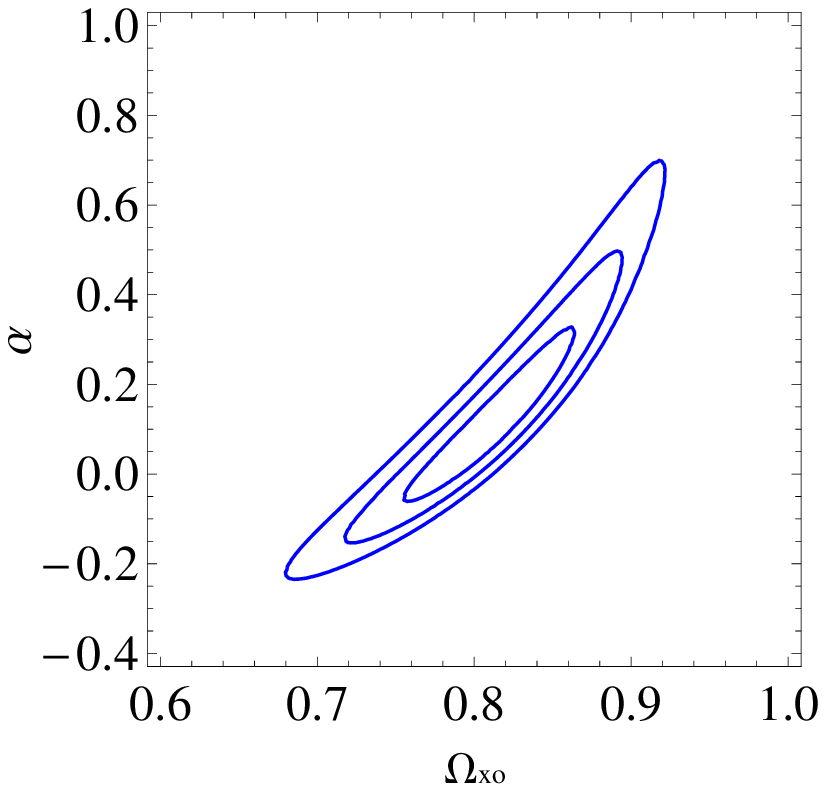}

\end{center}
\caption{In the top left we show the confidence regions for the Constitution dataset and in the top right for the Union dataset. In the bottom left confidence regions for the case of Constitution + BAO and in the bottom right for the case Union + BAO. In all cases we used $B$ given by Table 1 and a prior of $\Omega_{b0}=0.042$.}
\label{fig1}
\end{figure}

\begin{figure}[htb]
\begin{center}
\includegraphics[height= 7.0 cm,width=7.5cm]{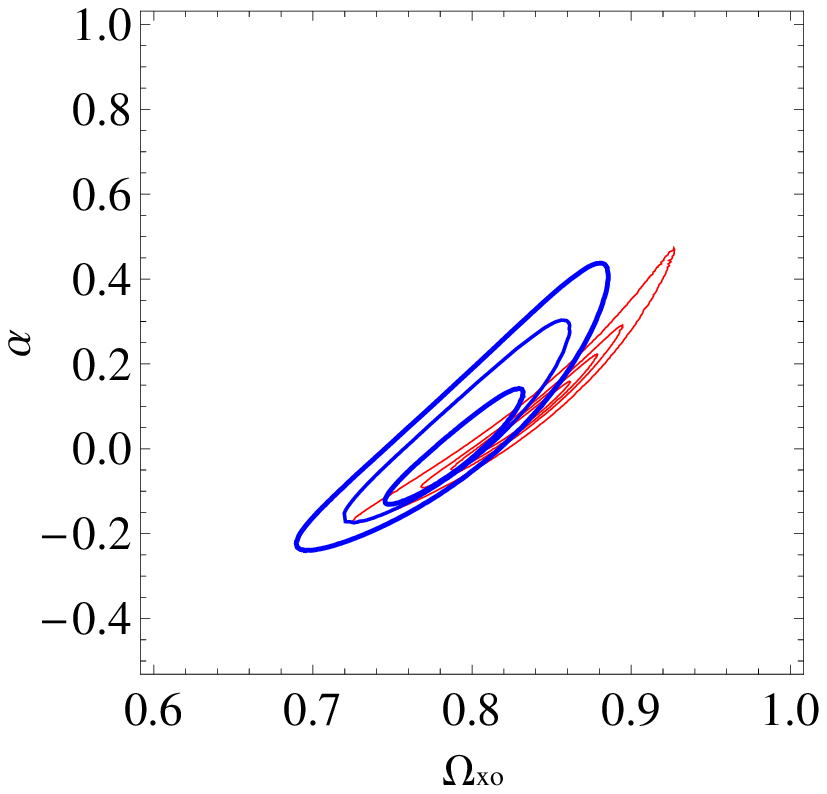}
\includegraphics[height= 7.0 cm,width=7.5cm]{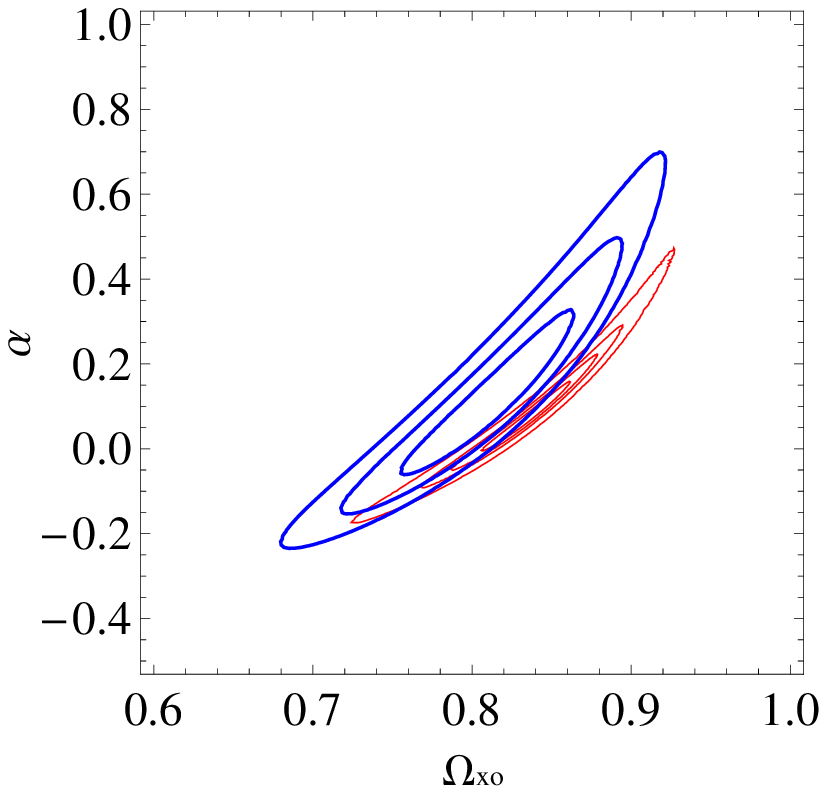}
\includegraphics[height= 7.0 cm,width=7.5cm]{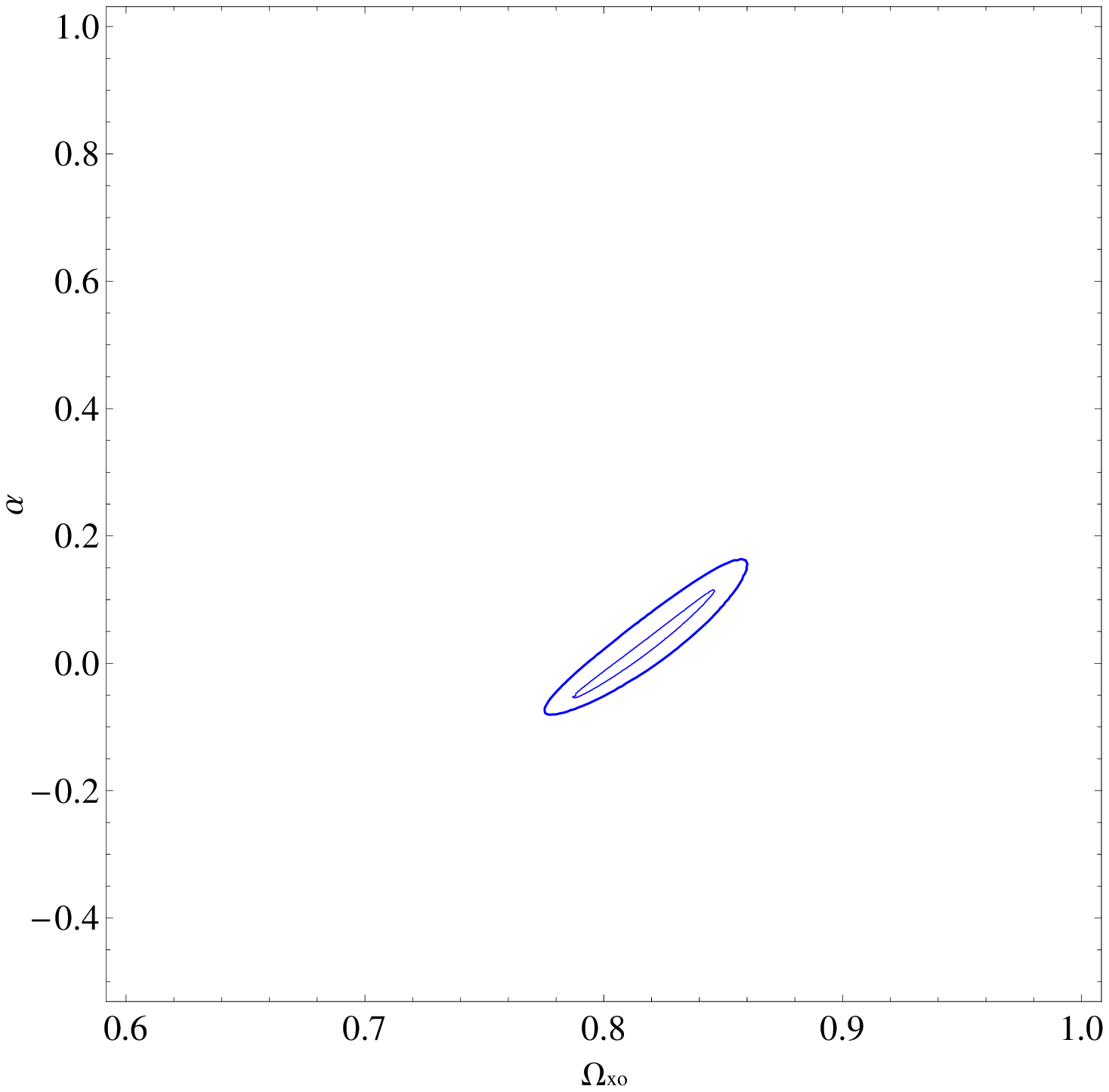}
\includegraphics[height= 7.0 cm,width=7.5cm]{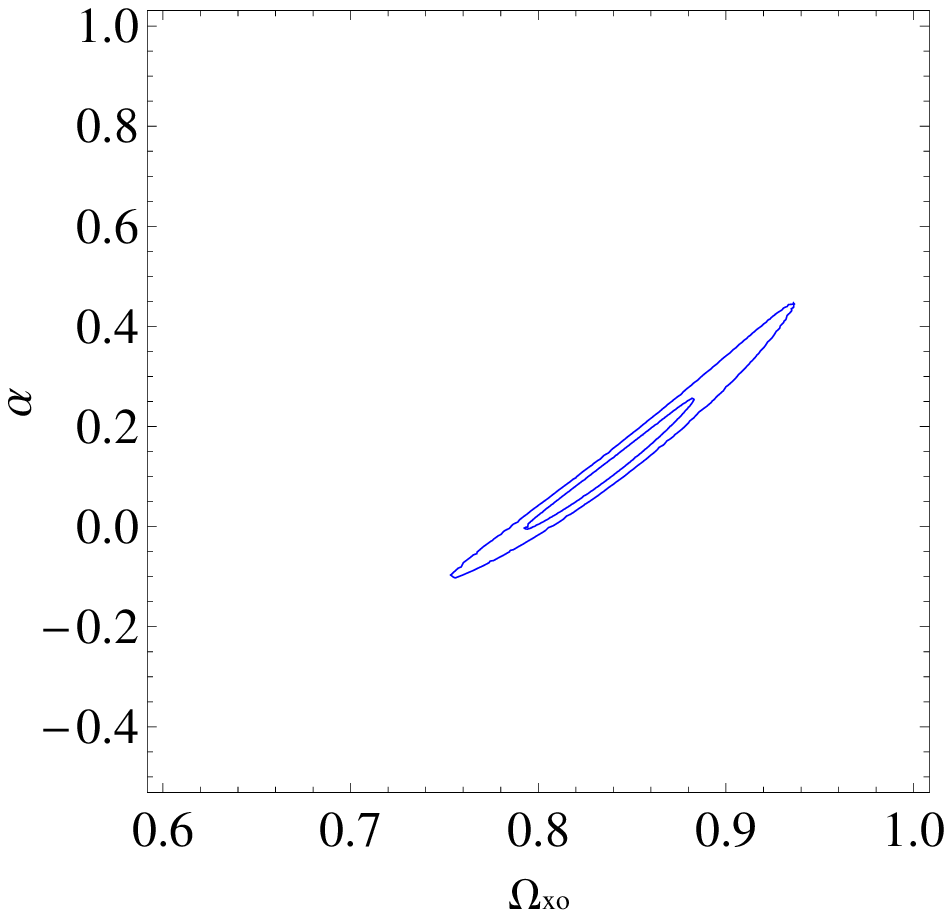}
\end{center}
\caption{In the top left we show observational constraints of SNeIa Constitution + BAO + CMB. In the top right, SNeIa Union + BAO + CMB. Each bottom panel shows the intersection of the curves above.}
\label{fig1}
\end{figure}

\end{document}